\documentclass[a4paper,fleqn]{cas-sc}
\usepackage{amsmath}
 \usepackage[numbers]{natbib}
\def\tsc#1{\csdef{#1}{\textsc{\lowercase{#1}}\xspace}}
\tsc{WGM}
\tsc{QE}
\tsc{EP}
\tsc{PMS}
\tsc{BEC}
\tsc{DE}
\ExplSyntaxOn \cs_gset:Npn \__first_footerline: { \group_begin: \small \sffamily \__short_authors: \group_end: } \ExplSyntaxOff

\begin{document}
\let\WriteBookmarks\relax
\def\floatpagepagefraction{1}
\def\textpagefraction{.001}

\shorttitle{Effect of peer influence factor and individual needs on social media contagion}

%\shortauthors{CV Radhakrishnan et~al.}

\title [mode = title]{Sustainability \& Social Segmentation in Social Media Contagion: A Mathematical and Computational Study on Dual Effects of Individual Needs \& Peer Influence}                      
%\tnotemark[1,2]


%\tnotetext[2]{}

\author[1]{Dibyajyoti Mallick}[]


%\fnmark[]

\ead{dm.20ph1103@phd.nitdgp.ac.in}

%\ead[url]{www.cvr.cc, cvr@sayahna.org}

\credit{Formal analysis, Software, Writing - Original draft preparation}

\affiliation[1]{organization={National Institute of Technology Durgapur},
    addressline={M.G Road, A Zone }, 
    city={Durgapur},
    % citysep={}, % Uncomment if no comma needed between city and postcode
    postcode={713209}, 
    country={India}}

\author[2]{Priya Chakraborty}
\credit{Methodology, Formal analysis, Visualization, Writing - Original draft preparation}
\author[1]{Sayantari Ghosh}
 \cormark[1]
%\fnmark[]
\ead{sayantari.ghosh@phy.nitdgp.ac.in}
%\ead[URL]{}

\credit{Resource, Supervision, Writing - Review}

\affiliation[2]{organization={Indian Institute of Technology Bombay},
     addressline={Powai}, 
    city={Mumbai},
    % citysep={}, % Uncomment if no comma needed between city and postcode
    postcode={400076}, 
    state={Maharashtra},
    country={India}}

\cortext[cor1]{corresponding author}
%\cortext[cor2]{Principal corresponding author}

%\fntext[fn1]{}
%\fntext[fn2]{}


\begin{abstract}
Addiction to internet-based social media has increasingly emerged as a critical social problem, especially among young adults and teenagers. Based on multiple research studies, excessive usage of social media may have detrimental psychological and physical impacts. In this study, we are going to explore mathematically the dynamics of social media addiction behaviour and explore the determinants of compulsive use of social media from the dual perspectives of individual needs or \textit{cravings} and peer-related factors or \textit{peer pressure}. The theoretical analysis of the model without the peer pressure effect reveals that the associated addiction-free equilibrium is globally stable whenever a certain threshold, known as the addictive-generation number, is less than unity and unstable when the threshold is greater than unity. We observed how introduction of peer influence adds a sustainability to the dynamics, and causes a multistability, through which addiction-contagion can proliferate, even below the designated critical threshold. Using simulations over model networks, we demonstrate our finding, even in the presence of social heterogeneity. Finally, we use the reaction-diffusion approach to investigate spatio-temporal dynamics in a synthetic society, in the form of a 2D lattice. Instead of a fast convergence to the steady states, we observe a long transient of social clustering and segmentation, represented by spatio-temporal pattern formation. Our model illustrates how the peer influence factor plays a crucial role and concludes that it is required to consider the peer factors while formulating specific strategies that could be more effective against this addiction and its potential adverse outcomes.
\end{abstract}

% \includegraphics{figs/grabs.pdf}

\begin{comment}
\begin{highlights}
\item A mathematical and simulation-based study on social media addiction, influenced by peers.
\item Peer-related factors had a stronger effect on individuals than their personal craving driven relapse, which has been verified using deterministic as well as network simulations.
\item Social segmentation in form of spatio-temporal clustering has been observed which represents a visual proliferation of social media addiction.
\end{highlights}
\end{comment}
% Keywords

\maketitle{}

\section{Introduction}
Presently, with the exponential growth of internet usage, there is an intense and compulsive urge, especially among the youth, to utilize social media (SM) platforms \cite{clement2020daily,tutgun2015development}, which frequently leads to severe overuse. Starting with the need to access information and curious attention, SM websites like Google, Twitter, Facebook, Instagram, and YouTube cause people to hinge on the platforms for various reasons, like finding old or new friends, earning money, making advertising products, buy or sell goods, make a money transaction, play games, going through funny memes and reels, etc. \cite{qualman2012socialnomics,gastaldi2014integration,mallick2024quantitative}. Attachment to one or more social media websites can be regarded as Social Media Addiction (SMA) when people spend excessive time on this and feel anxious when they cannot make a visit to that social media platform \cite{hou2019social}. % The World Health Organization (WHO) published a review report in 2019 that addressed the possible drawbacks of excessive social media use \cite{world2019guideline}.
Recent surveys have estimated that $4$ billion persons globally were active social media users. More than $73\%$ and $68\%$, respectively, of adult Americans are drawn to Facebook and YouTube \cite{smith2018social}. A recent cross-sectional survey has been done by Moreno et al. measuring problematic internet use, internet gaming disorder, different video gaming addiction, and social media addiction in young adults\cite{moreno2022measuring}. It has been reported that excessive social media use can also expose users to unfavorable emotions like distrust, depression, anxiety, low self-esteem, cyberbullying, triggering content, unhealthy social comparisons, sleep issues, melancholy, etc. \cite{ali2024mathematical,csenturk2021social}. This addiction habit especially affects a lot to the young generations, which leads to lower performance at work and poorer academic results \cite{pantic2012association}. For instance, studies in the US and Central Serbia have shown a positive correlation between young adults' depressive symptoms and the amount of time they spend on social media \cite{simic2016prevalence,muflih2018effect}. \\
It is interesting to note that most of the addictions, including SMA, are usually initiated by peer effects \cite{huang2021peer,lee2001assessment}. It has been reported that SM users, especially young users, get habituated to these platforms due to the frequent encouragement from their friends \cite{arikan2022two,mahamid2019social,steers2016influence}. Peer pressure is an interesting collective phenomenon and refers to the pressure that one feels when they are directly or indirectly asked to think and act according to the rules or requirements of their peers. It prioritizes group harmony over individual expression of thought and hence, peer pressure may be an essential predictor of an individual's social media addiction \cite{brzozowski2009effects}. \textcolor{black}{Thus, spreading in a peer group from person to person, SMA can be referred to as a social contagion \cite{ojiakor2024impact}, and epidemic models provide a powerful platform for quantitative exploration of this dynamics. Though designed initially to study the prevention of infectious disease dynamics \cite{siettos2013mathematical,hoan2020new}, in recent times, mathematical compartmental modeling has been used successfully to study social contagion phenomena. The contagious disease dynamics model has been used by many researchers to study the spreading of rumors, fake news, and addictions, like alcohol, drugs, internet, and gaming as well \cite{nasti2018computational,sanchez2007drinking}. These models typically rely on the concept of contagions arising through interpersonal interactions while the entire population is divided into a certain number of compartments. Taking the SIR model \cite{kermack1927contribution} as the backbone, several researchers have studied contagions of messages, habits, memes, campaigns, and opinions \cite{woo2011sir,gaurav2017equilibria,bhattacharya2019viral,blackmore2000meme}, and looked for interesting phenomena (like, bifurcations) with significant physical implications. Thus, as it follows similar dynamics, SMA can also be explored with this perspective.} \\
In this paper, we have developed a mathematical model depicting the social media addiction dynamics using the epidemiological modeling technique. We investigate how peer influence has a great impact on social media addiction and compare it with individual needs or cravings for the addiction. Considering how peer influences negatively sway addiction behavior, we consider withdrawal and then subsequent relapse of addiction habit in our model \cite{bahridah2023relationship}. In our model, while the introduction to the addiction is driven by a peer, relapse of the addiction after initial withdrawal can happen due to individual needs as well as peer pressure. Constructing a model that highlights peer-driven addiction as well as dual possibilities of relapse, we analyze SMA with deterministic analyses first. We look for inherent features like basic reproduction number and underlying bifurcations\cite{alexander2005bifurcation}. %Despite that, lots of work has been done based on this peer influencing factor, but it has not been well explored mathematically \cite{xu2023peer,wang2023examining}. Hence, in this paper, both mathematical and computational analysis are explored regarding this social media addiction dynamics. considering the linear transition once where the temporarily inert population who are not rigid about their decision are becoming addicted users again on their own or self cravings like they think that they should share on social media to communicate with whom and what they care about and psychological impact must have a significant influence here. considering this transition only we derive the basic reproduction number, the epidemic threshold determining whether this addiction will stay or vanish.  However, we get a saddle-node bifurcation when we have the nonlinear transition, including a peer-influenced relapse rate, and see even if the threshold is less than one, there are still addictive people \cite{alemneh2021mathematical}. That indicates Peer pressure had a more substantial effect on individuals than self-esteem. 
Next, we extend the model beyond the deterministic case to take into account society's heterogeneity and use complex networks to ascertain the dynamics of a synthetic society. Moreover, considering a reaction-diffusion counterpart of the proposed model, we explore the dynamics with visual understanding. %With some recent studies of Turing Pattern analysis for rumor dynamics systems, 
To look for spontaneous heterogeneities and social segmentation, we explore the temporal dynamics for SMA considering the spatial distribution of addictive behavior\cite{mallick2023visual} and study pattern formation considering the diffusion in a two-dimensional randomized population. We have organized the manuscript as follows: the respective compartmental model of social media addiction is proposed in Section \ref{Model formulation}. We have analyzed the bifurcations and thresholds mathematically through a deterministic approach \cite{ccakan2020dynamic,sharma2015backward} in Sections \ref{threshold} and \ref{bifurcations}. Afterward, we analyzed the numerical simulation of the model on a complex network in Section \ref{network}. We have next investigated pattern development for randomized populations while considering diffusion in a two-dimensional environment as discussed in Section \ref{pattern}. Finally, we conclude with a brief discussion along with some possibilities of future work in Section \ref{discussion}.

\begin{figure}
    \centering
    \includegraphics[scale=.4]{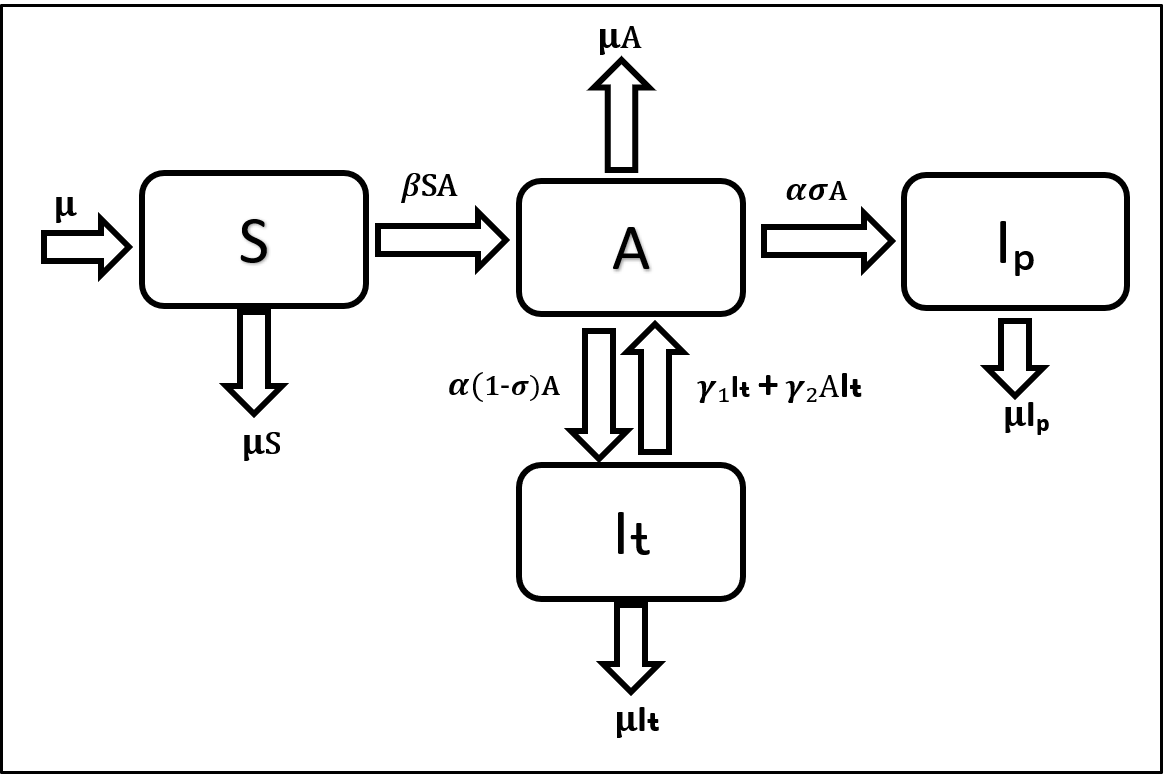}
    \caption{Block diagram of the proposed model of social media addiction. The notation for compartments (denoted by solid boxes) and transition rates (denoted by arrows with parameters) have been elaborated in the text.}
    \label{model}
\end{figure}
\section{Proposed Model}{\label{Model formulation}}
In this section, we consider a deterministic mathematical model for social media addiction (SMA) based on the assumption that the epidemic happens in a closed and homogeneously mixed population consisting of $N$ individuals. The dynamical process of the proposed $SAI_tI_p$ SMA spreading model is shown in Fig. \ref{model}, where the population is divided into four groups: Susceptible ($S$), Addictive ($A$), Temporary Inert ($I_t$), and Permanent Inert ($I_p$). The rules of the $SAI_tI_p$  model can be summarized as follows:
\begin{itemize}
    \item The Susceptible person becomes addictive with probability $\beta$, or spreading rate, upon contacting an addictive person. 
    In this group, individuals are not addicted yet to the use of any social media applications. Every individual in this class is denoted as $S$. They can acquire addiction habits from social media-addicted peers over time and most likely spend an unjustified amount of time online. Thus, per unit of time, the amount $\beta$$SA$ will be deducted from this population class and added to the rate equation of $A$. 
 \item The individuals who spend the majority of their time on social media and are addicted to it are portrayed here as addictive people and are denoted by $A$. As mentioned in the previous case, $\beta$$SA$ must be added to this class due to the increased number of addicted persons. Here, $\alpha$ is the rate by which addicted people are becoming Inert by quitting the use of social media due to the development of temporary boredom or awareness regarding SMA. So, the $\alpha$$A$ term must be deducted from this class and added to $I_t$. At the same time, those who are not rigid about their decision can again join this class on their own with rate $\gamma_1$ and be influenced by their peers with $\gamma_2$. So $\gamma_1$$I_t$ and $\gamma_2$$AI_t$ amount must be added here and deducted from $I_t$. Here the negative peer influence on relapse has been considered.     
  \item  People in the Temporary Inert group got over their SMA temporarily and have developed some ability to control their online actions. Each person in this class is denoted by $I_t$. Considering that only a fraction of $\sigma$ is rigid about their decision to leave SMA. So $\alpha(1-\sigma)$ $A$, people are designated as \textcolor{black}{Temporary Inert. They may not be rigid about their decision, and they revert to the addicted state because of either (i) their self-interest or craving or (ii) getting influenced by their addicted peers. The corresponding rates are considered as $\gamma_1$ and $\gamma_2$ respectively. As discussed in possible changes in class $A$, the respective rate should be subtracted from this class.}
  \item  People in the Permanent Inert group got over their SMA and have grown some ability to control their online actions permanently.  Each person in this class is denoted by $I_p$. Out of the total flux of $\alpha$ $A$ that leaves population $A$, a fraction of $\sigma$ remains in this Permanent Inert class. Thus, $\alpha$$\sigma$$A$ should be added to this class equation. 
\end{itemize}

For this analysis, We assume that the entire population is normalized to $1$:
\begin{equation}
     S(t)+A(t)+I_t(t)+I_p(t) = 1
 \end{equation}
 
Moreover, in this population, we consider $\mu$ as the rate at which people enter the dynamic system and leave, keeping the population size intact at any point in time so that we can maintain a variable demography. Thus, we have:

\begin{eqnarray}
\frac{dS}{dt}=\mu - \beta\;S\;A - \mu\;S  \nonumber\\ 
 \frac{dA}{dt}= \beta\;S\;A+(\gamma_1\ + \gamma_2\;A)\;I_t-(\mu\ + \alpha)\;A \\ 
\frac{dI_t}{dt}= \alpha\ (1-\sigma\ )A -(\gamma_1\ + \gamma_2\;A)\;I_t-\mu I_t  \nonumber\\
\frac{dI_p}{dt}= \alpha\ \sigma\;A-\mu I_p \nonumber
\end{eqnarray}

\begin{table}[width=.9\linewidth,cols=2,pos=h]
\caption{ Parameters used for the proposed $SAI_tI_p$ model and their physical meaning}\label{tbl1}
\begin{tabular*}{\tblwidth}{@{} LLLL@{} }
\toprule
Parameters Symbol & Description of parameters  \\
\midrule
$\mu$ &  Birth/death rate to maintain demography   \\
$\beta$ & Rate of addiction  \\
$\alpha$ & Rate of leaving the SMA   \\
$\gamma_1$ & Self-craving driven relapse rate  \\
$\gamma_2$ & Peer influenced relapse rate   \\
$\sigma$ & Fraction of addictives who permanently left the addiction    \\
\bottomrule
\end{tabular*}
\end{table}

\begin{figure}
    \centering
    \includegraphics[scale=.5]{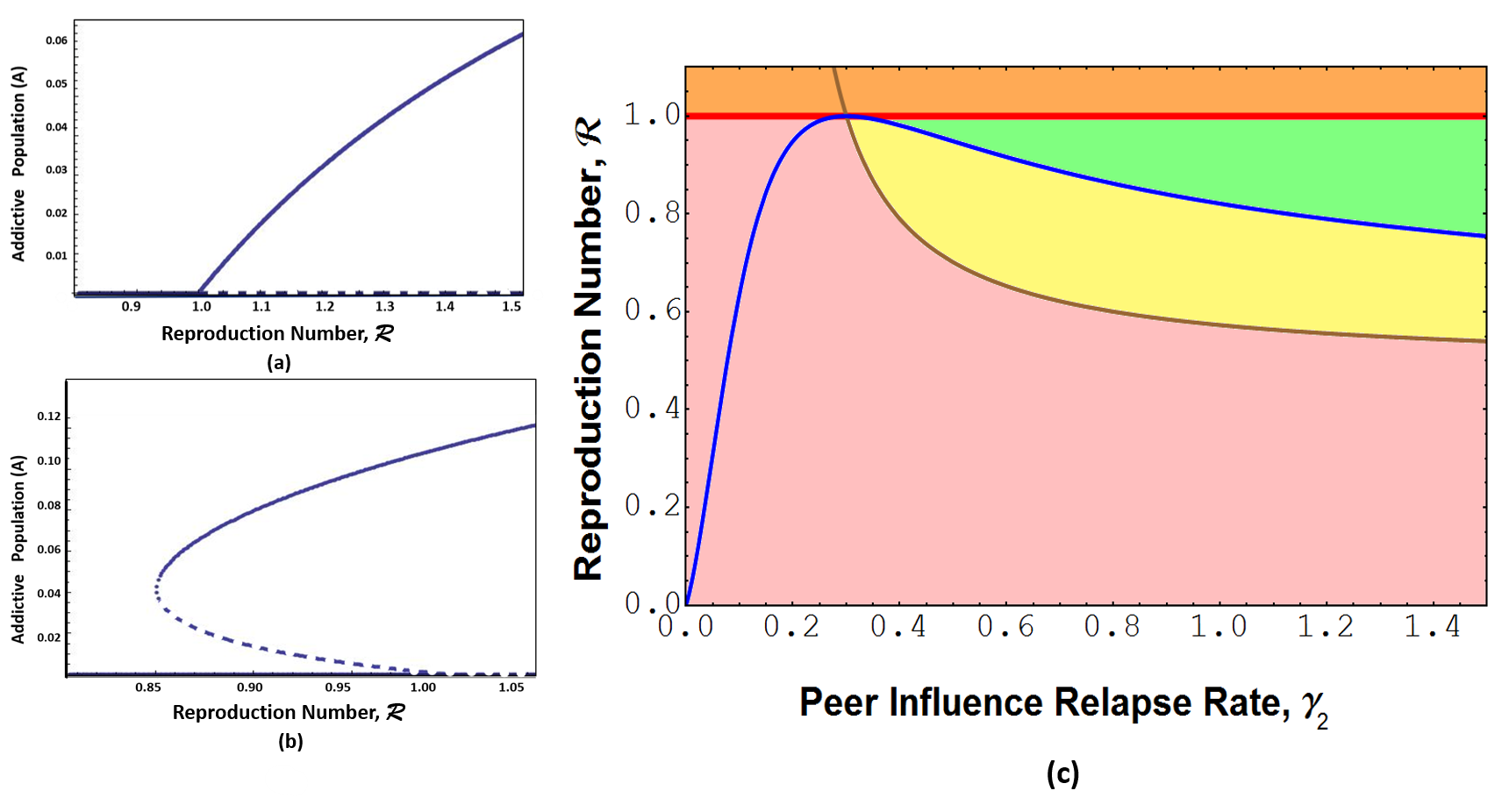}
    \caption{\textcolor{black}{Variation in steady state fraction of Addictive population with reproduction number $R$ for (a) $\gamma_2 = 0.015$, when only a single epidemic state persists beyond $R = 1$ and for (b) $\gamma_2 = 0.85$, when bistability can be observed in the range $R_c$ to $1$. In this figure, the continuous lines indicate stable solutions, and the dashed lines indicate unstable solutions. For these parameter values, we calculated $R_c = 0.85$  from equ \ref{critical threshold}. (c) Phase diagram of the model in $\gamma_2 - R$space for $\sigma = 0.2$ and $\mu = 0.01$. The blue line indicates $R_c$, the purple line indicates $\gamma_{2th}$, and the red line indicates $R = 1$. The region filled with orange color always exhibits a monostable endemic state as $R > 1$, and the pink region exhibits a monostable addiction-free state as $\gamma_2 < \gamma_{2th}$. For both the yellow and green regions, $\gamma_2 > \gamma_{2th}$. For the yellow region, $R < R_c$, and the region contains a monostable addiction-free state. However, for the green region, $\gamma_2 > \gamma_{2th}$, $R > R_c$, and $R < 1$. Thus, this area exhibits bistability, where either the addiction-free state or the endemic state is chosen by the system depending upon the initial state.}}
    \label{bif}
\end{figure}

\section{Basic Reproduction Number}{\label{threshold}}

In the absence of addiction, the model has an addiction-free equilibrium ($E_0$), which is given by
$E_0$ : $(S, A, I_t, I_p)$ = $(1, 0, 0,0)$.
The linear stability of the equilibrium  $E_0$ can be established in terms of the basic reproduction number ($R$): a potential quantity that determines whether the addiction can invade the population. Hence ($R$) is computed using the next-generation matrix technique proposed in \cite{van2002reproduction}. Thus, using the next-generation method right-hand sides of the equations for the rate of change of addiction present variables $\frac{dA}{dt}$ and $\frac{dI_t}{dt}$ must therefore be written in terms of two matrices, $F$ and $V$, where $F$ is a matrix made up of the addiction-generation terms and $V$ is an M-matrix made up of the remaining transition terms in the two equations.

That is the 2nd and 3rd rate equations of Eq. (2) are re-written as\\
\begin{equation}
  \begin{pmatrix}
\frac {dA}{dt} \\
\frac {dI_t}{dt} 
\end{pmatrix} 
 =
 \begin{Bmatrix}
     
 \begin{pmatrix}
\beta & $0$\\
$0$ & $0$
\end{pmatrix}
- 
\begin{pmatrix}
K_1 & -\gamma_1 \\
-K_2 &   K_3
\end{pmatrix}
\end{Bmatrix}
\begin{pmatrix}
$A(t)$\\
I_t(t)
\end{pmatrix} = (F-V) 
\begin{pmatrix}
$A(t)$\\
I_t(t)
\end{pmatrix},  
\end{equation}\\

with, $K_1$= $\mu$ + $\alpha$ , $K_2$= $\alpha$ (1- $\sigma$), $K_3$= $\mu$ + $\gamma_1$.\\
That is, for the model equation (2), the next-generation matrices given by 
\begin{equation}
  F= \begin{pmatrix}
\beta & $0$\\
$0$ & $0$
\end{pmatrix} and 
 V= \begin{pmatrix}
K_1 & -\gamma_1 \\
-K_2 &   K_3
\end{pmatrix}  
\end{equation}

Using the next-generation matrix the stability of $E_0$ is based on whether or not $\rho$($FV^{-1}$) $<1$, where $\rho$ is the spectral radius. And the basic reproduction number $R$ is defined as the spectral radius $\rho$($FV^{-1}$), which is the largest absolute value of the eigenvalues of $FV^{-1}$. If $\rho$($FV^{-1}$) $<1$, then all eigenvalues of the linearized system at $E_0$ have negative real parts, implying that $E_0$ is locally stable, and addiction cannot invade. For $\rho$($FV^{-1}$) $>1$, at least one of the eigenvalues of the linearization has a positive real part; thus, the $E_0$ is unstable in this case. Letting $R$ = $\rho$($FV^{-1}$), it is easy to show that \\
\begin{equation}\label{reproduction number}
R = \dfrac{\beta\ K_3}{K_1 K_3-\gamma_1K_2} = \dfrac{\beta\ (\mu+\gamma_1)}{\mu(\mu+\gamma_1)+\alpha (\gamma_1\sigma+\mu)}  = \frac{\beta}{\mu+\alpha- \frac{\alpha\;(1-\sigma)\;\gamma_1\;}{\mu+\gamma_1}}\\
\end{equation}
Hence for the system equations, the $E_0$ is locally asymptotically stable if $R <1$ and unstable if  $R >1$. \textcolor{black}{Interestingly, the reproduction number of the conventional SIR model: $R= \frac{\beta}{\mu+\alpha}$, if we consider similar notations for relevant parameters \cite{van2017reproduction}. In our model, the form of the reproduction number $R$ is found to be a function of 
%\begin{equation*}
 %   R = \frac{\beta}{\mu+\alpha\;\frac{\gamma_1\;\sigma+\mu}{\mu+\gamma_1}} = \frac{\beta}{\mu+\alpha+ \alpha\;\gamma_1\;\frac{\sigma-1}{\mu+\gamma_1}}
%\end{equation*}
the temporary withdrawal of the addiction behavior by self-relapse rate $\gamma_1$, the permanent fraction of addiction recovery rate $\sigma$, along with the rate parameters of conventional $SIR$ model. In general, $\sigma<1$ as it signifies the fraction of the population leaving the additive population permanently, thus in the denominator, the term $ \frac{\alpha(1-\sigma)\;\gamma_1}{\mu+\gamma_1}$ is always positive, and this increases the reproduction number from its conventional $SIR$ consideration. In other words, the temporary inert population in the dynamics, those who are not confident about their decision to leave social media addiction, rescales the reproduction number by the transition rate constants associated with $I_t$ compartment. For $\sigma=1$, our model converges to a conventional $SIR$ system as the rate of transferring to the temporary inert population goes to zero. This drops the consideration of peer factor and feedback in the dynamics, and the reproduction number also converges to the form of a conventional $SIR$ model reproduction number.
\\Physically, $R$ measures the average number of new addicts, generated from a single addicted person. The local stability result implies that the total number of addicted persons in the population can be reduced to zero if the initial sizes of the sub-populations of the model are in the basin of attraction of $E_0$. That is, a small influx of addicted people into the community will not generate a large number of addicts if $R <1$. Thus, an increase in the value of $R$ (compared to the conventional $SIR$ model), by the addition of the temporary inert population $I_t$ in our model, signifies an increase in the average number of new addicts generated from a single addicted person. Here, it is important to point out that the reproduction number $R$ is not a function of $\gamma_2$, the peer-influenced relapse rate, as shown in Eq. \ref{reproduction number}}. %the threshold of reproduction number $R_c$, in representing the coexistence of the bistable response of addicted and disease-free equilibrium in the system, is regulated by the peer influence rate of the system $\gamma_2$. 
Thus, we perform the bifurcation analyses in the presence of the nonlinear relapse rate to see the effect of peer factors and to derive the threshold conditions, in terms of parameter values in the next section. %To ensure that the effective control (or elimination) of the number of addicted persons in the community at steady-state is independent of the initial sizes of the sub-populations of the model, it is imperative to show that the AFE is globally asymptotically stable.

\begin{figure}
    \centering
    \includegraphics[scale=0.6]{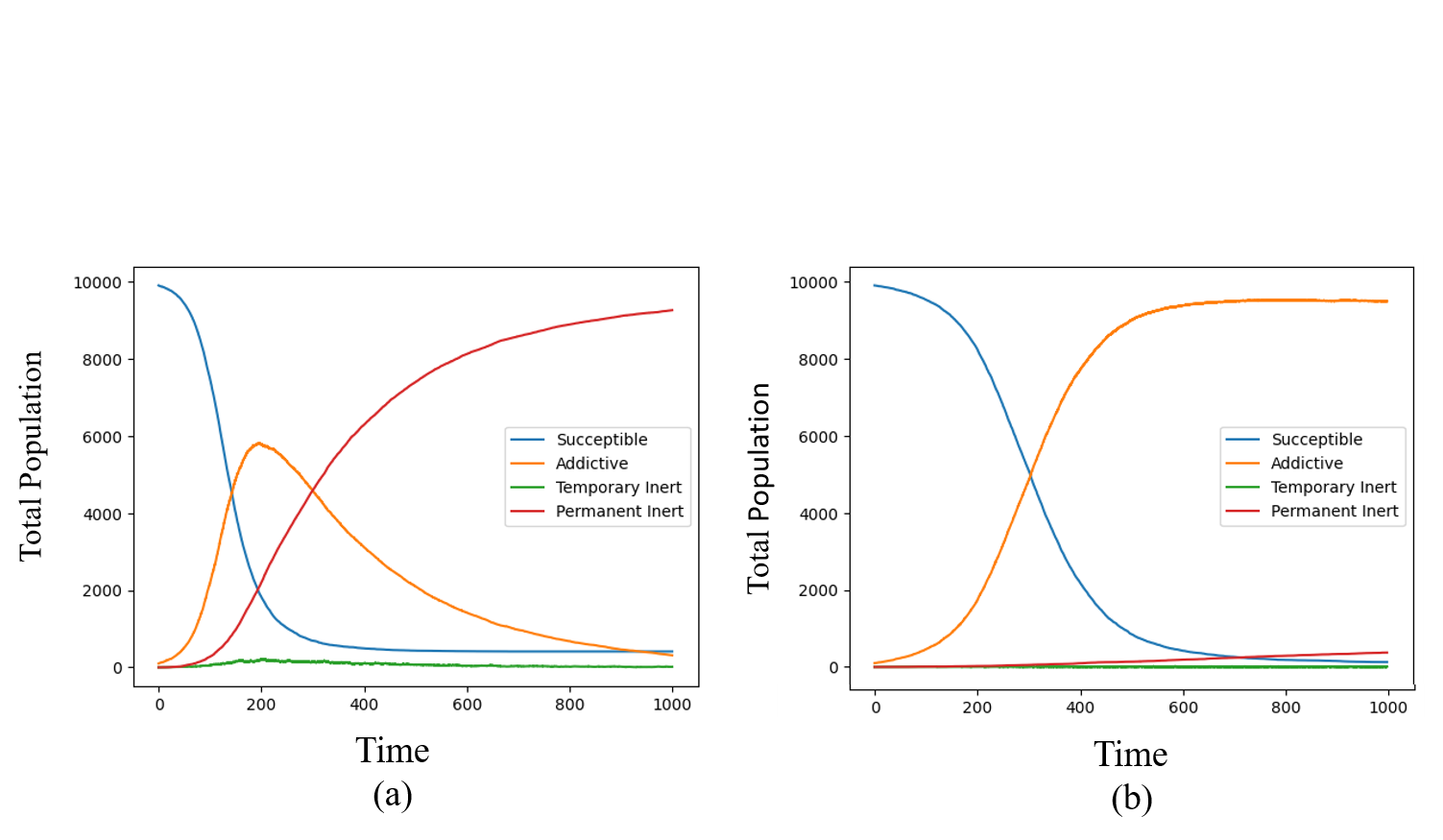}
    \caption{Time evolution of all the subpopulations for the proposed $SAI_tI_p$ model. (a) For $\gamma_2 = 0.015$ (less peer effect) (b) $\gamma_2 = 0.85$ (Strong peer effect). The rest of the parameter values for both curves are $\mu = 0.01$, $\beta = 0.0542$, $\sigma = 0.2$ and $\alpha = 0.01$}
    \label{rp on network}
\end{figure}
\section{Bifurcation Analysis}{\label{bifurcations}}

To understand the phenomenon in terms of both individual and peer effects, we consider the self-interest first (i.e., craving for social media making them again join the addiction) while there is a very little peer influenced relapse.  We observed a forward trans-critical bifurcation occurring at $R=1$ %for $\gamma_1 = 0.3$, 
indicating the existence of the addiction-free equilibrium only when this addictive-generation number is beyond unity, as shown in Fig. \ref{bif}(a). \\
On the other hand, while considering the nonlinear relapse rate, peer influence has a significant impact on the Addictive population to get back the addiction to social media. We got a backward (saddle-node) bifurcation at $R=1$ for a comparatively higher nonlinear relapse %$\gamma_2 = 0.85$, 
as shown in Fig. \ref{bif}(b). In this case, the system poses three solutions, out of which two are physically achievable, and the third one is an unstable fixed point. Out of the physically achievable solutions, one is $E_0$, the addiction-free solution, while the other is $E_1$, the addiction-endemic solution. For this parameter regime where $R\in (R_c, 1)$, both these solutions exist, is known as the region of bistability, where $R_c<1$. This depicts the possibility of getting addictive endemic state, even below $R=1$, and marks a regime where both the social media-addicted or temporarily inert population can dominate the population, depending upon the initial conditions. It is widely recognized that these bistable dynamics introduce memory into the system, causing it to behave history-dependently. Specifically, while the system is in the bistable zone, it remembers its prior state and does not go to the other steady state for slight fluctuations in the parameter value. It can be stated that, if the backward bifurcation is present in the dynamics, %while $R<1$, the system chooses to stay in its high $A$ value for the inherent non-linearity present in the dynamics, and 
it becomes much more difficult to eradicate the addiction from the system, because the peer influencing factor is higher. The condition of existence of this bistable region will be discussed in \ref{bistability} 

\subsection {\textcolor{black}{Conditions for bistability}}{\label{bistability}}
\textcolor{black}{The model system ($2$) always has an Addiction-free equilibrium $E_0$$(1, 0, 0,0)$, at which the whole population is not addicted.
Also, this system exhibits an Endemic equilibrium $E_1$$(S^\star, A^\star, I_t^\star, I_p^\star)$.
At equilibrium, equating the rate of change in different population compartments to zero,
 the first equation of the system model becomes 
\begin{equation}
  S^\star = \frac{\mu}{\beta A+\mu}   
\end{equation}}

 \textcolor{black}{Relevant substitutions from Eq. (6) and replacing $I_t$ by $(1-S-A-I_p)$, simple algebra results into $pA^2 + qA + r = 0$,
 where 
 \begin{eqnarray}
       p = \alpha\beta\sigma\gamma_2 + \mu\beta\gamma_2  \nonumber\\ 
       q = \alpha\beta\sigma\gamma_1 + \beta\mu\gamma_1 +\alpha\mu\sigma\gamma_2 + \mu^2\gamma_2 + \mu^2\beta + \mu\alpha\beta - \mu\beta\gamma_2 \\
       r= \mu\gamma_1\alpha\sigma + \mu^2\gamma_1 + \mu^3 + \mu^2\alpha -\mu^2\beta - \mu\beta\gamma_1 \nonumber
 \end{eqnarray}}

\textcolor{black}{To ensure bistability, the necessary conditions are $q>0$ and $q^2 - 4pr > 0$ where $p$, $q$, and $r$ are given by Eq. (7). These relations allow us to determine the limiting condition for bistability. The nonlinear relapse rate, $\gamma_2$ causes a drastic change in the behavior of the system, though it does not appear in the expression of $R$. We can figure out the minimum threshold for $\gamma_2$ (say $\gamma_{2th}$) by equating $q=0$ as follows: 
 \begin{equation}
  \gamma_{2th} = \frac{\alpha\sigma\gamma_1\beta+ \gamma_1\mu\beta+\mu^2\beta+\mu\alpha\beta}{\mu\beta-\mu^2-\alpha\sigma\mu}
  \label{gamma2thres}
 \end{equation}}

\textcolor{black}{For a given set of parameters, if $\gamma_2$ > $\gamma_{2th}$, then bistable solutions can be expected. Once we satisfy the condition for $\gamma_2$, it is essential to note that both the endemic states can exist (\i.e., have real roots) only if $q^2 - 4pr > 0$. The region of bistability extends for a range of $R$ values, from $R_c$ to 1, as discussed earlier. For $R < R_c$, the endemic solutions are not feasible, and the only steady state is addiction-free. $R_c$, or the critical threshold for bistability, can be evaluated by equating $q^2-4pr$ to zero. With algebraic manipulations, we first identified the critical $\beta$ (say $\beta_c$) value, and furthermore, we have solved for the critical reproduction number (say $R_c$).}

\textcolor{black}{\begin{equation}
    \beta_c = \frac{\gamma_2^2\mu^2(\mu+\alpha\sigma)^2}{\gamma_2\mu(\mu+\alpha\sigma)(\mu(\alpha+\gamma_1+\gamma_2+\mu)+\alpha\gamma_1\sigma)-2\sqrt{\alpha\gamma_2^3\mu^4(1-\sigma)(\mu+\alpha\sigma)^2}}
\end{equation}}

\textcolor{black}{\begin{equation}
    R_c = \frac{\beta_c(\mu+\gamma_1)}{(\mu(\mu+\gamma_1)+\alpha(\mu+\sigma\gamma_1)}
    \label{critical threshold}
\end{equation}}
\textcolor{black}{Eqns. \ref{gamma2thres} and \ref{critical threshold} set two critical limits for achieving bistable behavior in the system. From our previous discussions, it is evident that depending upon the two key parameters of the model, $R$ and $\gamma_2$, the system can reach one of three outcomes: an endemic state only, an addiction-free state only, or both states simultaneously. To illustrate this idea, we explored the phase diagram of the system in $\gamma_2 - R$ space in \ref{bif}(c). Only the area shaded in green, which is enclosed by the boundaries: Eqn. \ref{gamma2thres}, Eqn. \ref{critical threshold} and $R = 1$, exhibits bistable dynamics. Fig \ref{bif}(a) and (b) can be obtained by tracking the system behavior while moving across the phase space through a vertical line $\gamma_2 = 0.85$ (and $\gamma_2 = 0.015$), as shown in Figs. 2(a) and 2(b). We note that the region of bistability increases gradually as the value of $\gamma_2$ increases. This shows that high values of the relapse rate ensure the survival of the addiction in a steady state. From the phase diagram, it can also be noted that without the relapse ($\gamma_2 = 0$), no bistability is possible.}

\begin{figure}
    \centering
    \includegraphics[scale=0.5]{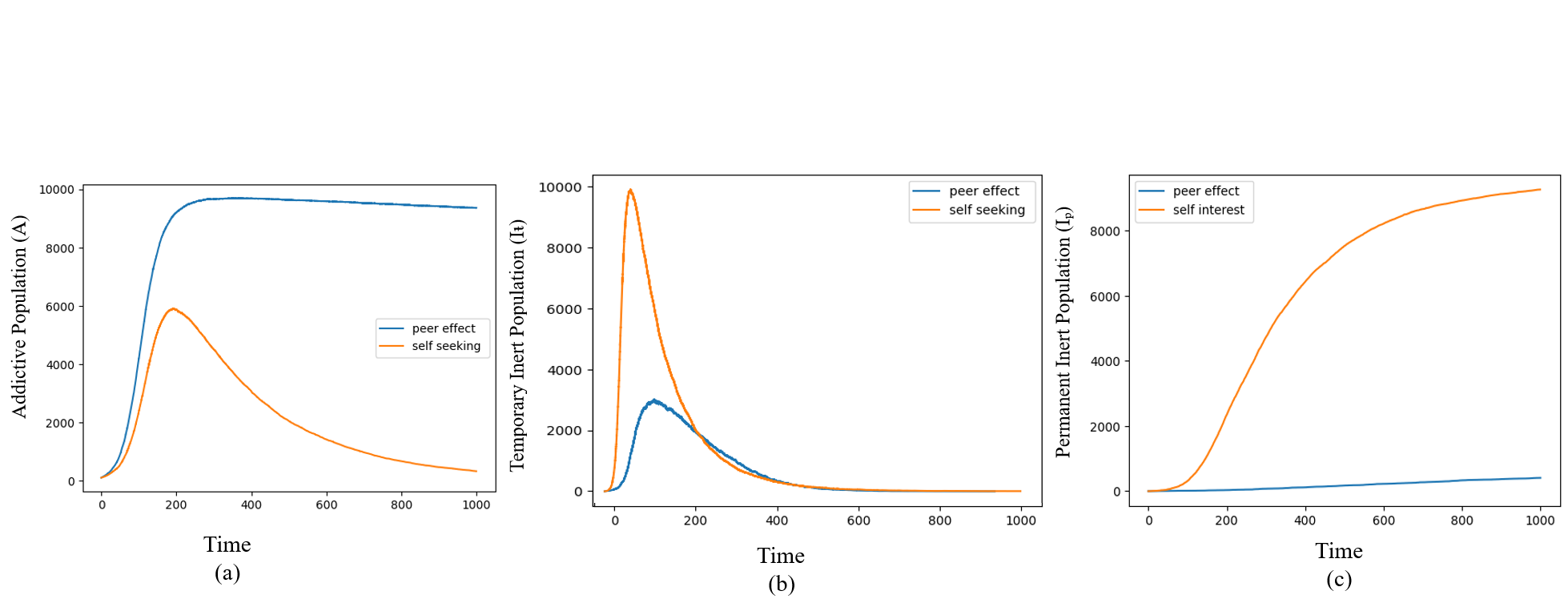}
    \caption{Population of different compartments with time for Individual craving vs. peer effects towards SMA (a) Addictive Population (b) Temporary Inert Population (c) Permanent Inert Population. Here for self-seeking $\gamma_1 = 0.015$ and $\gamma_2 = 0$ and for peer effects $\gamma_1 = 0$ and $\gamma_2 = 0.85$. The rest of the parameter values are the same as Fig. \ref{rp on network}.}
    \label{peer on network}
\end{figure}

\section{Network Analysis}{\label{network}}

In this section, we report the numerical simulation result of our proposed model over the network. One of the main problems with ODE-based models is homogeneous mixing, which assumes that each member of a population has an equal chance of interacting with every other member of the society. Hence, considering the fact that our society is quite diverse we inspect the model in the complex network's heterogeneous environment.  We consider a network of $N$ individuals as a social network, where individuals are nodes and contacts between people are edges. Then, a graph $G = (V, E)$  can be obtained, where $V$ is the set of nodes and $E$ is the set of edges. We assume that social media addiction is disseminated by direct and indirect contacts of addictive nodes with others.\\ 
The simulation is conducted on a random network having 10000 nodes with an average degree of 5. Here, we have used the EoN module in Networkx on Python to run the simulation in Google Colaboratory. Based on the deterministic analysis discussed in the preceding section, \ref{bifurcations} shows that the peer effect relapse parameter $\gamma_2$ is responsible for this negative influence. As we all know, peer influence occurs when someone is inspired or persuaded to do something by seeing their friends or family. \textcolor{black}{People are often gets influenced negatively by their peers to engage in harmful activities}, and that is spotted here in terms of the steady state of existence of addictive people in Fig. \ref{rp on network}(b), where the peer influence is much stronger compared to Fig. \ref{rp on network}(a).  We can see from Fig.\ref{rp on network}(b) where, $\gamma_2 = 0.85$, the impact of peers leads to a growth in the addictive population that finally saturates over time at a non-zero high value of addicted people. On the other hand, for $\gamma_2 = 0.015$, when there is less peer factor, it increases and attains a peak and then saturates to a much lower value as shown in Fig. \ref{rp on network}(a). It happens because the system exhibits bistability. An exciting finding here is even though in deterministic analysis when there is no peer effect, the steady state value related to $A$ population was \textit{zero}. Here, because of the diversity of the synthetic society, some addictive individuals will always persist, even in a steady state (Fig. \ref{rp on network}(a)). This clarifies that our society, as a network here, is so diverse and highly heterogeneous, and certain individuals can never truly recover from addiction and become harmful peer influencers.\\
In  Fig.\ref{peer on network} we have compared how all the populations behave with time, with the presence of peer effect vs. driven by their self-cravings towards the use of different social media platforms. The parameters have been chosen such that the orange graphs (marked as self-seeking in Fig. \ref{peer on network}) refer to $\gamma_1 = 0.015, \gamma_2 = 0$, while the blue curves (marked as peer effect) refer to $\gamma_1 = 0, \gamma_2 = 0.85$. In  Fig.\ref{peer on network}(a), As people are negatively affected by their peers hence, the addiction stays in society for a long time while the addiction eventually fades up in case of self-seeking. Temporary Inert and Permanent Inert population evolves accordingly, as shown in Fig. \ref{peer on network}(b) and Fig.\ref{peer on network}(c). We clearly note the drastic difference, where most of the population is in an addictive state for the blue set (peer effect present), while most are in a permanent inert state for the orange set (peer effect is nearly absent). It signifies that peer influence is an essential factor in this kind of addiction spreading, and peer pressure has a more substantial effect on individuals, than self-interest.

\section{Spatiotemporal Analysis}{\label{pattern}}
\subsection{Model formulation for spatiotemporal analysis} To understand the spatiotemporal behavior of the spreading process of addiction in the population, we now consider a two-dimensional lattice of population. In a $200 \times 200$ array of population, the addiction behaviour is considered to be diffused. Two different diffusion constants, $D_A$ and $D_I$, have been considered for diffusion of addictive behaviour in the Addicted and the Temporary Inert population respectively, where $D_A > D_I$, considering the temporary detachment of $I_t$ population from SMA. Here, we consider isotropic diffusion (the rate of diffusion/ diffusion coefficient is the same in both directions of the considered two-dimensional population lattice) and explore the dynamics in a no-flux boundary condition. In the presence of diffusion, the system dynamics can be represented by the following set of partial differential equations (PDE):

\begin{eqnarray}
\frac{\partial S (x,y,t)}{\partial t}=\mu - \beta\;S\;A - \mu\;S \\ \nonumber
 \frac{\partial A (x,y,t)}{\partial t}= \beta\;S\;A+(\gamma_1\ + \gamma_2\;A)\;I_t-(\mu\ + \alpha)\;A + D_A\;\nabla^2 A\\ \nonumber
\frac{\partial I_t (x,y,t)}{\partial t}= \alpha\ (1-\sigma\ )A -(\gamma_1\ + \gamma_2\;A)\;I_t-\mu I_t + D_I\;\nabla^2 I\\
\frac{\partial I_p (x,y,t)}{\partial t}= \alpha\ \sigma\;A-\mu I_p\nonumber
\end{eqnarray}
where the $\nabla^2$ is the Laplacian operator in $\mathbb{R}^n.$ Now in order to maintain the sum of all population equal to $1$, we set the initial condition as
\textcolor{black}{
\begin{eqnarray}\label{initial}
 S(x,y,0) = k_1\\ \nonumber
 A(x,y,0) = k_2\;\zeta (0,1)\\ \nonumber
 I_t(x,y,0) = 0\\ \nonumber
 I_p(x,y,0)=1- S(x,y,0)-A(x,y,0) 
\end{eqnarray}}
Here, $k_1$ and $k_2$ are two scaling factors taken as $k_1 =0.9$ and $k_2=0.1$, such that at $t=0$, the sum of the population becomes $1$. A randomization term $\zeta$ is considered that picks up a number between (0,1) in order to initialize the addiction in the population dynamics. With this setup, we further proceed to explore the spatio-temporal dynamics of the system.
\begin{figure}
    \centering
    \includegraphics[width=0.85\textwidth]{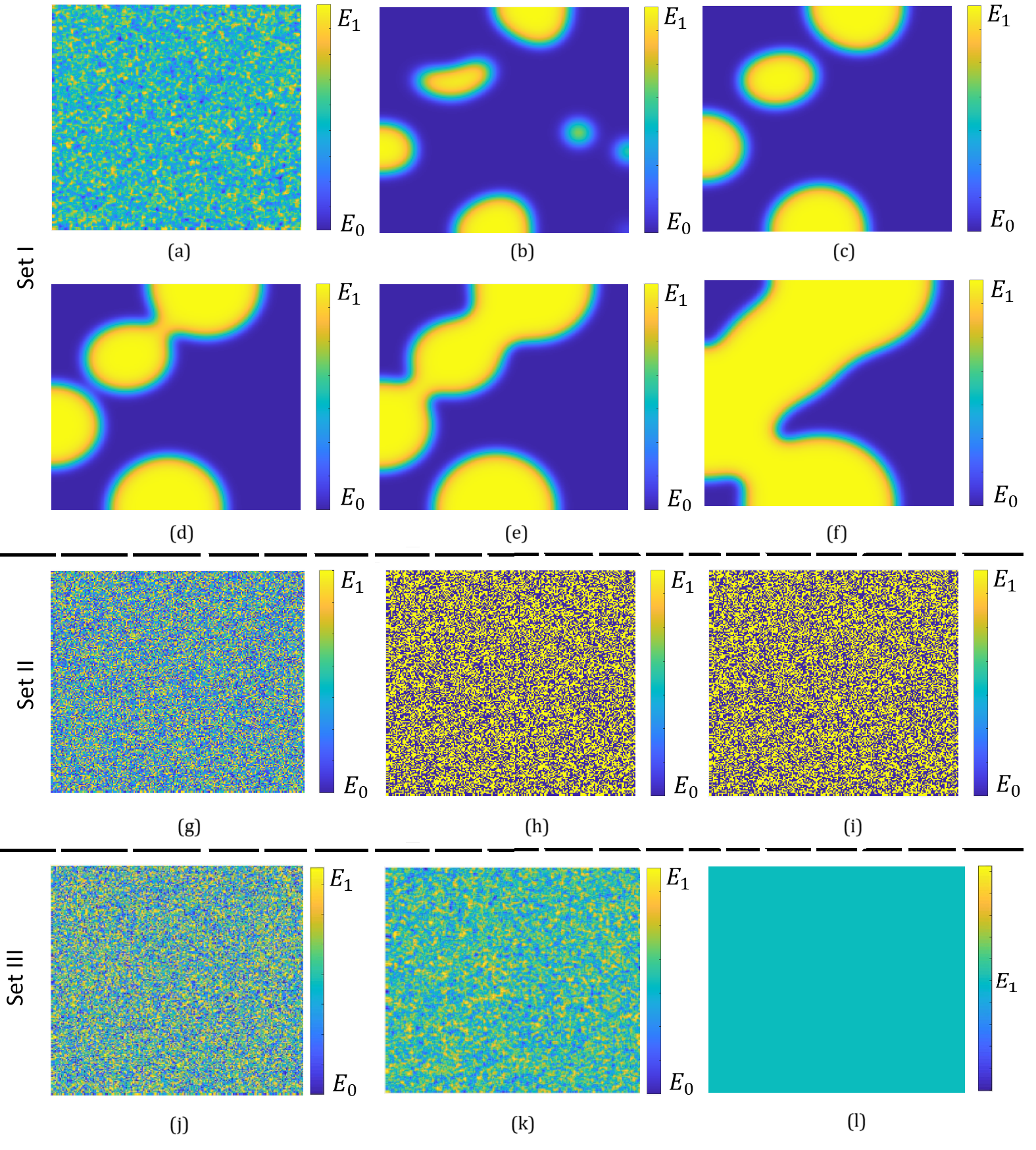}
    \caption{\textcolor{black}{The effect of diffusion in the spatiotemporal behavior of the addictive population under peer effect. Set I: The system shows cluster formation in the presence of diffusion of the addicted population to reach the endemic state under peer effect. Yellow islands represent endemic clusters, and blue islands represent disease-free equilibrium clusters. Set II: In the absence of diffusion of addiction, the system spatially shows bistability with time; however cluster formation is not observed. Parameter values are $\mu=0.01, \gamma_1 = 0.015,\gamma_2 = 0.85,\sigma = 0.2, \beta = 0.0542, D_a =0.05, D_I = 0.0015$.  for (a) to (i). For (a) to (f) $D_A= 0.05,D_I=0.0015$, for (g) to (i) $D_A= 0,D_I= 0$. Set III: (j)-(l) Under negligible peer effect, in the presence of diffusion no cluster formation is observed. $\gamma_1=0.15$, $\beta=0.0642$ for (j) to (l), the rest of the parameters are the same as Set I.  Snapshots are taken after time Set I: (a) 10 (b) 5000 (c) 9450 (d) 12750 (e) 14200 (f) 19350 Set II: (g) 10 (h) 11000 (i) 15000 Set III: (j) 10 (k) 200 (l) 2000. } }
    \label{addicted pattern}
\end{figure}
\subsection{Social Segmentation driven by Bistability \& Diffusion} 
\begin{figure}
    \centering
    \includegraphics[width=\textwidth]{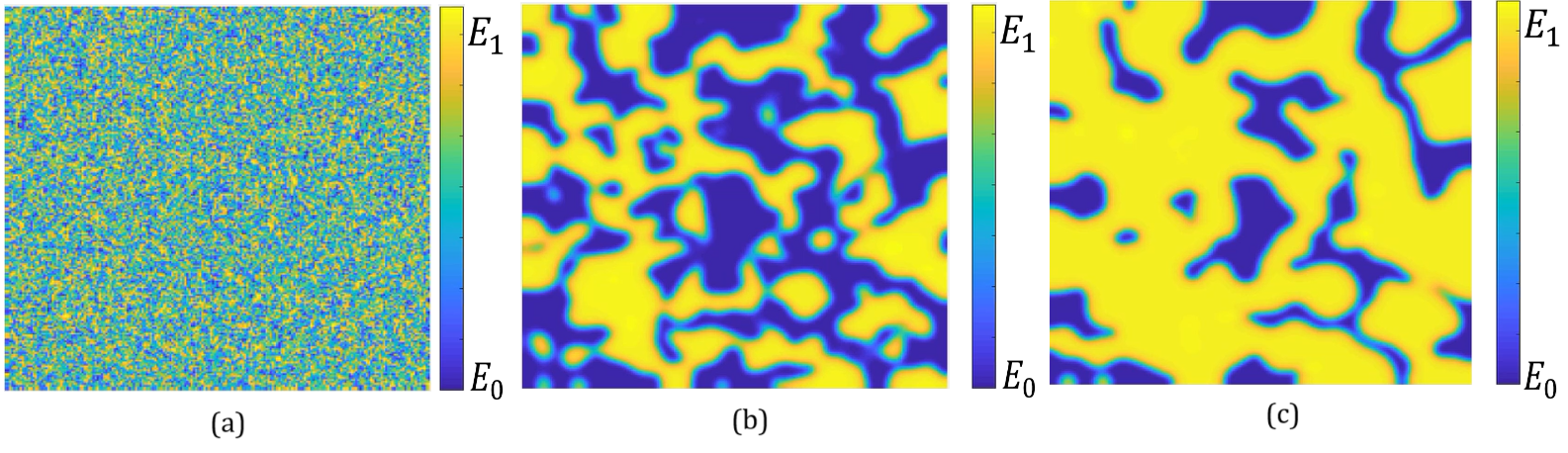}
    \caption{Under the peer effect, with a diffusion of the addictive behavior, the system shows cluster formation in order to reach the endemic state. Spatiotemporal evolution of permanent inert population ($I_p$) dynamics. The initial condition is given by  $S(x,y,0) = k_1; 
 A(x,y,0) = k_2\;\zeta (0,1);
 I_t(x,y,0) = 1- S(x,y,0)-A(x,y,0);
 I_p(x,y,0)=0$. Yellow islands represent endemic clusters, and blue islands are representative of disease-free equilibrium clusters. Snapshots are taken after the time (a) 30 (b) 1450 (c) 2250. Parameter values are $\mu = 0.01,  \sigma =0.15 $,  $\alpha = 0.055,\gamma_1=0,\gamma_2=0.3,\beta=0.047, k_1=0.9,k_2=0.1$.  Diffusion coefficient associated with addictive population $D_A= 0.02$ and for temporary inert population $D_{I}=0.0015$. }
    \label{spatio}
\end{figure}

\textcolor{black}{ The spatiotemporal dynamics for the addicted population are shown in Fig. \ref{addicted pattern}, for three different conditions. Our model system shows cluster formation under the diffusion of addiction behavior in the presence of peer influence.  The dynamics for the addicted population are shown in Fig. \ref{addicted pattern}(a)-(f). }
 %Starting from the randomized initial condition, as shown in Fig. \ref{addicted pattern}(a), with time, some clusters of $E_0$ and $E_1$ are observed in the output. With respect to time, the system will eventually reach its endemic state,  which is a straightforward outcome for addiction/disease dynamics,
 Two separate states ($E_0$ and $E_1$) coexist spatially for a prolonged intermediate time, forming evolving clusters of addicted populations spatially. The increase of yellow islands with time, while connecting smaller islands to larger ones (thus, blue islands are turning yellow), shows the effect of peer influence on a local environment of behavioral diffusion. \textcolor{black}{The clustering nature is regulated by the existent bistability and the diffusive addiction proliferation in the system. To further understand this, we simulated the same setup (Fig. \ref{addicted pattern} Set I), in the absence of diffusion (Fig. \ref{addicted pattern} Set II).  The randomized initialization perturbs the system, which evolves to its nearby steady states ($E_0$ and $E_1$) with time because of the underlying bistability of the system in the presence of high peer influence. However, no cluster formation is noticed in the long time limit. 
In the condition of low peer influence/absence of peer influence,  the spatiotemporal dynamics is straightforward and monotonically evolves to the steady state value with time representing an endemic condition. Fig. \ref{addicted pattern} Set III (j) to (l) shows the dynamics for different times, for $A$ population. Thus, broadly in terms of system parameters, under high peer influence, the dynamics show bistability, governed by the conditions mentioned in Sec. 4, and spatially in the presence of diffusion the dynamics show richness by formation of evolving clusters of endemic and disease-free population. The cluster formation is regulated by both the bistability and the diffusion of the addiction in the population lattice (Fig. \ref{addicted pattern} Set I), an absence of either (absence of diffusion in Fig. \ref{addicted pattern} Set II, absence of peer influence Fig. \ref{addicted pattern} Set III) spatially restricts the clustering nature in order to reach the endemic state. } \\
It is important to mention that we have shown the cluster formation for the $A$ population, to demonstrate the social segmentation in the addictive habit; however, as expected in coupled dynamics, similar clusters are observed for all other populations in the presence of peer influence in the diffusible population lattice, given the condition for bistability is satisfied. For example, temporal evolution of spatial pattern for the $I_p$ population for a different parameter set %and initial condition 
is shown in Fig. \ref{spatio}.%Next, let us proceed to quantify the size of the clusters with time.  
\textcolor{black}{\subsection{Connected Components \& Measure of Patch Size}}
%\begin{figure}
 %   \centering
 %   \includegraphics[width=0.85\textwidth]{3950 for da=0.02.png}
  %  \caption{(a)-(b)Pattern formation in addicted population. Snapshots are taken after time (a) 3950 (b) 10600. (c)-(d)Measure of cluster formation for the addicted population, averaged over 50 systems. Parameter values are $\mu=0.01, \gamma_1 = 0.015,\gamma_2 = 0.85,\sigma = 0.2, \beta = 0.0542, D_a =0.02, D_I = 0.0015$. }
   % \label{cluster}
%\end{figure}

\begin{figure}
    \centering
    \includegraphics[width=\textwidth]{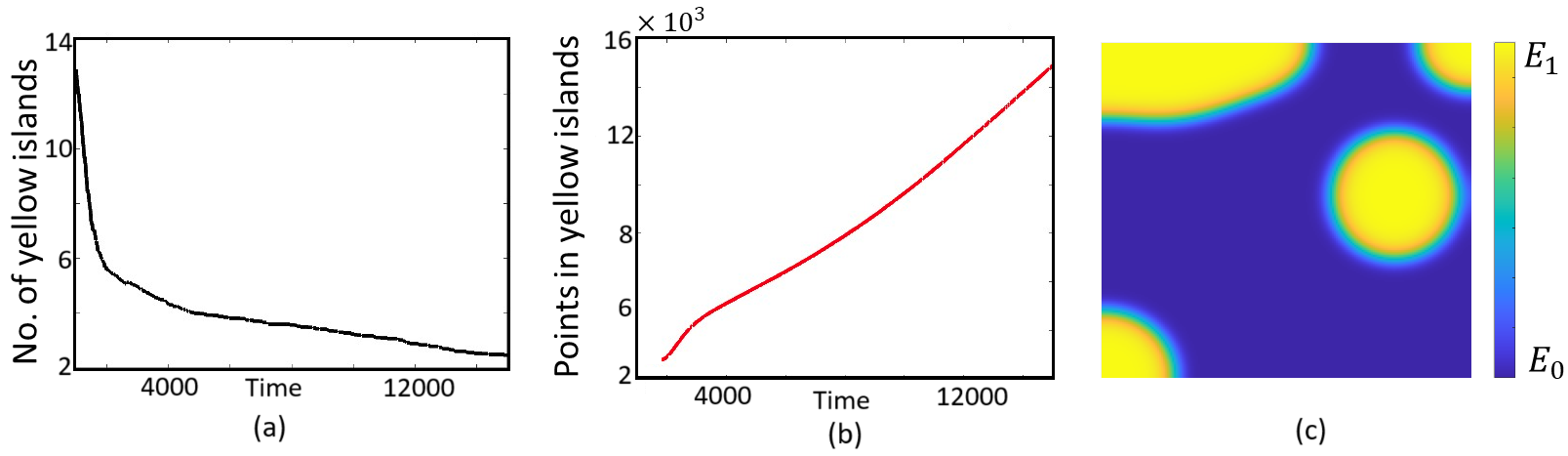}
    \caption{\textcolor{black}{Measure of cluster formation in the presence of peer influence, averaged over 50 realizations. With time we count the number of yellow islands (a), and the lattice points in yellow islands (b), for addiction diffusion coefficient $D_A = 0.05$. A visual of the clustered population state is shown in (c) after time $15000$. Other parameter values are $\mu=0.01, \gamma_1 = 0.015,\gamma_2 = 0.85,\sigma = 0.2, \beta = 0.0542, D_I = 0.0015$.}}
    \label{cluster}
\end{figure}

\textcolor{black}{    Diffusion in the considered population spreads the addiction from the local lattice to the neighboring global one.  We have quantitatively measured the spatiotemporal dynamics of the patterned proliferation of the addiction in our model system by connected component analysis on the discretized space. We consider the population lattice as the domain of the diffusion process $(x,y,t) \in \mathbb{Z} $ and perform a threshold operation to estimate $A_{th}$ such that $A_{th} = \frac{A^{max}+A^{min}}{2}$. Here, $A^{max}$ is the maximum possible value of $A \forall (x,y,t)$, and $A^{min}$ is the minimum of $A$. Next, we determine $\hat{A}(x,y,t)$ and considered it as one(zero) if $\hat{A}(x,y,t)$ is greater(less) than $A_{th}$. Considering the yellow points as $\hat{A}$ $= 1$ as the object points we perform the connected component analysis to find the number of the yellow island. We count the total yellow points at any time instant by counting the points where $\hat{A}$ $= 1$. We have averaged the system for $50$ realizations. % We performed this quantitative analysis for two different values of $D_A$ to understand the effect of diffusion on reaching the endemic solution by the system. In Fig. \ref{cluster} (a)-(c)  $D_A = 0.05$  in Fig. \ref{cluster} (d)-(f) keeping the rest of the parameters same.
    Overall the number of yellow islands decreases with time as the smaller islands join together to form a large cluster (Fig. \ref{cluster}(a)). Incidentally, the number of points in yellow islands increases with time as blue lattice points  ($E_0$) turn yellow ($E_1$) with time (Fig. \ref{cluster}(b)). %With the increase in diffusion coefficient, comparing Fig. \ref{cluster}(b) and \ref{cluster}(e) it is clear that the system shows a delay in reaching the endemic solution. Out of the $40\times10^3$ population lattice, the number of points in yellow islands, i.e. lattice point in endemic state $E_1$ is less in Fig. \ref{cluster}(e) than in \ref{cluster}(b) after an equal amount of time ($t=15000$) elapsed. 
    A visual of the clustered population, representing the same is shown in Fig. \ref{cluster}(c).% An increase in the diffusion of addictive nature in the population will sustain the instability for longer, and the system shows prolonged clustering.
    This quantitatively measures the trend in the spread of addiction in the dynamics via diffusion.}

\section{Conclusion}{\label{discussion}}

\textcolor{black}{Given the growing concerns about compulsive social media use, it is critical to comprehend why developing solutions to stop impulsive social media engagement is harder than expected. As reported by various studies, poor sleep quality, exhaustion, behavioral and lifestyle changes, social anxiety, decreased academic engagement in students are only some of the deep concerns associated with SMA. The findings of a recent study \cite{wang2022effects} demonstrated that peer relationships, school atmosphere, and parent-child proximity all had a substantial direct impact on teenagers' problematic Internet use. Thus, our goal is to investigate the elements that contribute to obsessive social media use from the viewpoints of both peer-related and personal demands. These phenomena demand closer examination through the lens of epidemic models to quantify, understand and address these contagions. Though models of epidemic dynamics are being used to study various social contagions, we found the reason due to which the standard, ‘simple’, contagion often fails to capture both the rapid spread and the sustainability associated with SMA problem. \\
We observed that the persistent nature of this dynamics is not only dictated by the initial spreading rate and reproduction number, but also affected by the difficulty to stick to the decision of withdrawal, due to personal cravings and peer influence. With elaborate mathematical treatment, we identify the role of both these factors. We find that the parameter representing self-driven individual needs effectively increase the reproduction number, contributing to faster spread of the contagion. On the other hand, peer influence has a more pressing effect, enhancing the nonlinearity and causing the existence of the endemic state, much below the expected threshold. Moreover, as we see, with strong peer influence, the addiction achieves a sustainability, driven by the inherent hysteresis of associated multistability. %In this paper, mathematical and computational analyses are explored regarding this social media addiction dynamics. In this study, with four compartments: Susceptible, Addicted, Temporary Inert, and Permanent Inert, we report how the peer influence factor frequently precedes personal self-interest and negatively affects getting back social media addiction. Using both mathematical and computational analysis, we consider a linear relapse transition (where the temporarily inert population, who are not rigid about their decision, are becoming addicted users again on their own or self-cravings) for which the parameter $\gamma_2$ is responsible, and a nonlinear relapse rate which is $\gamma_1$ (where peer pressure pushes them to the addicted state once again) as well. Considering the model, we derive the basic reproduction number, or the addiction threshold, determining whether this addiction will stay or vanish. 
We get a saddle-node bifurcation when we have the nonlinear transition, including a peer-influenced relapse rate, and see even if the threshold $R<1$, there are still addictive people in the steady state. The phase diagram of the dynamics, in light of the most important parameters $R$ and $\gamma_2$ gives a rich implication of the threshold values to ensure this multistability; it has been also shown that the region of bistability grows as the value of $\gamma_2$ the nonlinear relapse rate increases. We explained this through network analysis as well, as we have shown how negative peer influence affects people to stay in an addicted state in the steady state.   While we have observed that the steady states of dynamics in networks are closely related to deterministic system dynamics, we recognized the implications of structured contact networks in this context.\\
One of our major findings has been obtained when we observed this social system in a diffusive environment to study the structural proliferation of addictive behavior. We have noted interesting pattern formation in the presence of peer influence, which depicts a social clustering of addictive and aware/inert populations. Slow, long transients exhibiting social segmentation are observed in the form of discrete islands or \textit{endemic domains}. We have seen how peer influence plays an essential role in this depiction, rather than their self-motivation, as Fig. \ref{addicted pattern} shows no pattern formation in the absence of peer effect. The possible existence of bistability in this dynamics reflects the immense sustainability of the addictive state, and shows how difficult it could be to get rid of this addiction habit. %The results showed that peer pressure significantly predicted social media addiction. Hence, the importance of self-interest in comparison with the effect of peer influences could be eliminated. 
This result can have major implications in designing effective intervention strategies to ensure effective control (or elimination) of the number of addicted persons in the community, based on the understanding of the causes of the persistent nature of the contagion. It will be helpful to construct strategies like setting limits, minimizing screen time, and asking for help when necessary, which addresses overcoming peer pressure. It will be interesting to observe effect of positive peer-effect in this context. The aspects of network topology could also be taken into account for a better understanding of these kinds of dynamics, which may be further studied in future work.}

%\appendix

%\section{Bibliography}

% \appendix
% \section{My Appendix}
% Appendix sections are coded under \verb+\appendix+.

% \verb+\printcredits+ command is used after appendix sections to list 
% author credit taxonomy contribution roles tagged using \verb+\credit+ 
% in frontmatter.


\bibliographystyle{cas-model2-names}

\bibliography{cas-refs}


\end{document}